\documentstyle[12pt]{article}
\begin{document}
\begin{center}
{\Large \bf
Indeterministic Quantum Gravity} \\[0.5cm]
{\large\bf IV. The Cosmic-length Universe and the Problem of
the Missing Dark Matter} \\[1.5cm]
{\bf Vladimir S.~MASHKEVICH}\footnote {E-mail:
mashkevich@gluk.apc.org}  \\[1.4cm]
{\it Institute of Physics, National academy
of sciences of Ukraine \\
252028 Kiev, Ukraine} \\[1.4cm]
\vskip 1cm

{\large \bf Abstract}
\end{center}

This paper is a sequel of the series of papers [1-3], being
an immediate continuation and development of the latter of them.
In the Friedmann universe, the equation $\Omega_0=2q_0$ holds
($\Omega$ is density parameter, $q$ is deceleration parameter,
and subscript 0 indicates present-day values), which gives rise
to the problem of the missing matter as observational data give
$\Omega_0<2q_0$. In the cosmic-length universe, $\Omega_0=2q_0-
L/R_0^3H_0^2$ ($R_0$ is the radius of the universe, $H_0$ is
Hubble constant), which lifts the problem. The cosmic length,
$L=const \approx 1/H_0$, is the infimum of the set of maximal
radii of a closed universe.

\newpage

\hspace*{9 cm}
\begin{minipage}[b]{6 cm}
Let me have length...
\end{minipage}
\begin{flushright}
Ballad "Robin Hood's Death" \vspace*{0.8 cm}
\end{flushright}

\begin{flushleft}
\hspace*{0.5 cm} {\Large \bf Introduction}
\end{flushleft}

The most important problem facing modern cosmology is that
of the missing dark matter [4]. The problem has two aspects,
one relating to the universe as a whole, the second
concerning tiny nonuniformities in the temperature of a
remainder radiation. In this paper, we deal only with the
first aspect. Most of the mass of galaxies and an even larger
fraction of the mass of clusters of galaxies is dark. The
problem is that even more dark matter is required to account
for the rate of expansion of the universe.

More specifically, for the Friedmann universe, the equation
$\Omega_0=2q_0$ holds, where $\Omega=\rho/\rho_c$, $\rho$ is
density, $\rho_c$ is its critical value, $q$ is the deceleration
parameter, and subscript 0 indicates present-day values. In
particular, if $q_0>1/2$, the universe is closed and $\rho_0>
\rho_{c0}$. But observational data give $\Omega_0<2q_0$.

As is shown in [3], in cosmodynamics, which is based on
the theory being developed in this series of papers, for the
Robertson-Walker spacetime there is a possibility of a closed
universe with a density which is smaller than the critical
one. The present paper is an immediate continuation and
development of the paper [3]. The most important results are
as follows.

In place of $\Omega_0=2q_0$, the equation $\Omega_0=2q_0-
L/R_0^3H_0^2$ holds, where $R$ is the radius of the universe, $H$
is Hubble constant, and $L=const$ is an integral of motion
introduced in [3], which we call cosmic length. The model of
the universe with $L\ne 0$ we call the cosmic-length universe.
The Friedmann universe is that with $L=0$. The equation
$L=[(2q_0-\Omega_0)/(2q_0-1)]R_0$ holds. As $R_0\approx 1/H_0$,
we have $L\approx 1/H_0$ (but not $L=0$!). The equation $R_{\rm max}
=L+(8\pi\kappa/3)R_{\rm max}^3\rho|_{R=R_{\rm max}}$ takes place, where
$\kappa$ is the gravitational constant; thus
$R_{\rm max}\ge L$, in which sense the cosmic-length universe is
bounded below, as opposed to the Friedmann universe. The cosmic
length is the infimum of the set $\{R_{\rm max}:L\le R_{\rm max}<\infty
\}$.

For the cosmic-length universe as a whole, the problem of the
missing matter does not exist.

The cosmic length dwarfs Planck length: $L\approx 10^{26}\:{\rm m}>>>
1.6\cdot 10^{-35}\:{\rm m}=l_P$. These constants are measures of the
greatness and smallness respectively.

\section{The Robertson-Walker spacetime and observational
parameters}

The metric of the Robertson-Walker spacetime is of the form

$$g=dt^2-R^2(t)\left\{\frac{dr^2}{1-kr^2}+r^2d\theta^2+
r^2\sin^2\theta\,d\varphi^2\right\},\quad k=-1,0,1.
\eqno{(1.1)}$$

In models of the universe related to this spacetime, matter
is described by density, $\rho(t)$, and pressure, $p(t)$.
The models differ from one another by equations for $R$,
$\rho$, and $p$.

Observational parameters are Hubble constant
$$H_0=\frac{\dot R_0}{R_0}, \eqno{(1.2)}$$
deceleration parameter
$$q_0=-\frac{\ddot R_0R_0}{\dot R_0^2}, \eqno{(1.3)}$$
and density parameter
$$\Omega_0=\left(\frac{\rho}{\rho_c}\right)_0=\frac{\rho_0}
{3H_0^2/8\pi\kappa},\qquad \rho_c=\frac{3H^2}{8\pi\kappa},
\eqno{(1.4)}$$
where dot denotes $d/dt$, $\kappa$ is the gravitational
constant, and subscript 0 indicates present-day values.

\section{The Friedmann universe and the problem of the
missing dark matter}

The Friedmann universe is a model in which the complete
Einstein equation is used [5,3]. The latter amounts to
two equations:
$$2\ddot RR+\dot R^2+k=-8\pi\kappa pR^2, \eqno{(2.1)}$$
$$\dot R^2+k=\frac{8\pi\kappa}{3}\rho R^2. \eqno{(2.2)}$$
From eqs.(2.1),(2.2) it follows
$$\dot\rho R^3+3(\rho+p)R^2\dot R=0, \eqno{(2.3)}$$
which is equivalent to
$$dE=-pdV,\quad E=\rho V,\quad V\sim R^3. \eqno{(2.4)}$$

For present-day values, eqs.(2.1),(2.2) reduce to
$$p_0=-\frac{1}{8\pi\kappa}\left[\frac{k}{R_0^2}+H_0^2
(1-2q_0)\right], \eqno{(2.5)}$$
$$\rho_0=\frac{3}{8\pi\kappa}\left(\frac{k}{R_0^2}+
H_0^2\right), \eqno{(2.6)}$$
or
$$\Omega_0=1+\frac{k}{R_0^2H_0^2}. \eqno{(2.7)}$$

For the closed universe $(k=1)$, it follows from eq.(2.2) for
the maximal value of the radius
$$R_{\rm max}=\sqrt{\frac{3}{8\pi\kappa\rho}}, \eqno{(2.8)}$$
so that the infimum of the set of $R_{\rm max}{}'$s is equal to zero:
$$\inf\{R_{\rm max}:R_{\rm max}<\infty\}=0. \eqno{(2.9)}$$

As
$$p_0\ll\frac{1}{3}\rho_0, \eqno{(2.10)}$$
it follows from eqs.(2.5),(2.6)
$$\frac{k}{R_0^2}=(2q_0-1)H_0^2. \eqno{(2.11)}$$
Eqs.(2.7),(2.11) give
$$\Omega_0=2q_0. \eqno{(2.12)}$$

It is the equation (2.12) that is the primary source of the
problem of the missing dark matter: According to observational
data
$$\Omega_0<2q_0, \eqno{(2.13)}$$
or
$$\rho_0<\rho_{0\rm Friedmann}, \eqno{(2.14)}$$
where
$$\rho_{0\rm Friedmann}=\frac{3H_0^2q_0}{4\pi\kappa} \eqno{(2.15)}$$
is a value of $\rho_0$ given in the Friedmann universe. The
value
$$\Delta\rho_0=\rho_{0\rm Friedmann}-\rho_0 \eqno{(2.16)}$$
is the missing part of the density.

As long as
$$p\ll\frac{1}{3}\rho \eqno{(2.17)}$$
holds, the equality
$$\rho R^3=\rho_0R_0^3 \eqno{(2.18)}$$
is fulfilled. Then the equation of motion for $R$
$$\left(\frac{\dot R}{R_0}\right)^2=H_0^2\left[1-2q_0+2q_0
\left(\frac{R_0}{R}\right)\right] \eqno{(2.19)}$$
takes place [5].

\section{The cosmic-length universe and the lifting of the
problem}

In cosmodynamics [3], the dynamical equation (2.1) holds,
whereas the constraint (2.2) is not valid. The equation for
matter is (2.4). Thus in place of eqs.(2.1),(2.2), we have
equations
$$2\ddot RR+\dot R^2+k=-8\pi\kappa pR^2, \eqno{(3.1)}$$
$$\frac{d(\rho R^3)}{dR}=-3pR^2 \eqno{(3.2)}$$
((3.2) is equivalent to (2.4)). It is pertinent to note that
$p$, $\dot\rho$, and, as a consequence, $\ddot R$ undergo
sudden changes at quantum jumps of the state of matter.

As
$$2\ddot RR+\dot R^2=\frac{d}{dR}(R\dot R^2), \eqno{(3.3)}$$
we obtain from eqs.(3.1),(3.2)
$$\frac{d}{dR}\left( R\dot R^2+kR-\frac{8\pi\kappa}{3}\rho
R^3\right)=0, \eqno{(3.4)}$$
whence
$$R\dot R^2+kR-\frac{8\pi\kappa}{3}\rho R^3=L=const.
\eqno{(3.5)}$$

The length $L$---an integral of motion---we call cosmic length.
In accordance with this, the model considered is called the
cosmic-length universe.

It is seen from eqs.(2.2),(3.5) that the Friedmann universe
corresponds to a particular value of the cosmic length,
$$L_{\rm Friedmann}=0. \eqno{(3.6)}$$
In this sense, the Friedmann universe is the zero-length
universe.

The results for present-day values are as follows. From
eq.(3.1) we obtain
$$p_0=-\frac{1}{8\pi\kappa}\left[\frac{k}{R_0^2}+
H_0^2(1-2q_0)\right], \eqno{(3.7)}$$
i.e., eq.(2.5). From eq.(3.5) it follows
$$\rho_0=\frac{3}{8\pi\kappa}\left(\frac{k}{R_0^2}+H_0^2-
\frac{L}{R_0^3}\right), \eqno{(3.8)}$$
or
$$\Omega_0=1+\frac{k-L/R_0}{R_0^2H_0^2} \eqno{(3.9)}$$
in place of eqs.(2.6),(2.7) respectively.

As the inequality (2.10) holds, we obtain from eqs.(3.7),(3.8)
$$\frac{k}{R_0^2}=(2q_0-1)H_0^2, \eqno{(3.10)}$$
i.e., eq.(2.11). Eqs.(3.9),(3.10) give
$$\Omega_0=2q_0-\frac{L/R_0}{R_0^2H_0^2} \eqno{(3.11)}$$
in place of eq.(2.12).

Eqs.(3.11),(3.9) lift the problem of the missing matter for
the universe as a whole. In the cosmic-length universe this
problem does not exist.

In indeterministic quantum gravity---a theory being developed
in this series of papers---the universe is closed, i.e.,
$$k=1, \eqno{(3.12)}$$
so that by eq.(3.10)
$$q_0>\frac{1}{2} \eqno{(3.13)}$$
in agreement with observational data.

We proceed to consider the cosmic-length universe. From eqs.
(3.5),(2.18),(3.12) we obtain
$$R_{\rm max}=\frac{8\pi\kappa}{3}\rho_0R_0^3+L \eqno{(3.14)}$$
and, using eqs.(3.11),(1.4),
$$R_{\rm max}=2q_0H_0^2R_0^3, \eqno{(3.15)}$$
which is independent of $L$. From eqs.(3.11),(3.15) it
follows
$$L=\frac{2q_0-\Omega_0}{2q_0}R_{\rm max}. \eqno{(3.16)}$$
We obtain from eqs.(3.5),(2.18),(3.10),(3.11)
$$\left(\frac{\dot R}{R_0}\right)^2=H_0^2\left[1-2q_0+
2q_0\left(\frac{R_0}{R}\right)\right], \eqno{(3.17)}$$
i.e., eq.(2.19). It follows from
eqs.(3.10),(3.11),(3.16),(3.12)
$$R_0=\frac{1}{\sqrt{2q_0-1}H_0}, \eqno{(3.18)}$$
$$L=\frac{2q_0-\Omega_0}{2q_0-1}R_0, \eqno{(3.19)}$$
$$R_{\rm max}=\frac{2q_0}{2q_0-1}R_0. \eqno{(3.20)}$$
In the Friedmann universe, eqs.(3.18),(3.20) are fulfilled,
whereas eq.(3.19) reduces to the identity 0=0.

\section{The cosmic length as the infimum of the set of
maximal radii}

We have from eqs.(3.18),(3.19) for the cosmic length
$$L\approx R_0\approx\frac{1}{H_0}. \eqno{(4.1)}$$

It follows from eq.(3.5) with $k=1$
$$R_{\rm max}=L+\frac{8\pi\kappa}{3}(\rho R^3)_{R=R_{\rm max}}.
\eqno{(4.2)}$$
Thus, in the cosmic-length universe, the infimum of the set
of $R_{\rm max}{}'$s is equal to $L$:
$$L=\inf\{R_{\rm max}:R_{\rm max}<\infty\}. \eqno{(4.3)}$$

\section{What is small and what is great?}

The equations of the cosmic-length universe involve two
constants: the gravitational constant
$$\kappa=t_P^2, \eqno{(5.1)}$$
where $t_P$ is Planck time, and the cosmic length $L$. Now
that we have two universal constants---Planck length $l_P$
and the cosmic length $L$---we may answer the question:
What is small and what is great? It is these constants that
are measures of the smallness and greatness respectively.
And the greatness dwarfs the smallness:
$$L\approx 10^{26}\:{\rm m}>>>1.6\cdot10^{-35}\:{\rm m}=l_P. \eqno{(5.2)}$$
The universe is great in the sense of this inequality.


\begin{thebibliography}{9}

\bibitem{1} Vladimir S. Mashkevich, {\it Indeterministic
Quantum Gravity} (gr-qc/9409010, 1994).

\bibitem{2} Vladimir S. Mashkevich, {\it Indeterministic
Quantum Gravity II. Refinements and Developments}
(gr-qc/9505034, 1995).

\bibitem{3} Vladimir S. Mashkevich, {\it Indeterministic
Quantum Gravity III. Gravidynamics versus Geometrodynamics:
Revision of the Einstein Equation} (gr-qc/9603022, 1996).

\bibitem{4} Steven Weinberg, {\it Dreams of a Final Theory}
(Vintage, London etc., 1993).

\bibitem{5} Steven Weinberg, {\it Gravitation and Cosmology}
(John Wiley and Sons, Inc., New York etc., 1972).
\end{thebibliography}
\end{document}